\documentstyle[sprocl]{article}

\def\Journal#1#2#3#4{{#1} {\bf #2}, #3 (#4)}

\def\PLB{{\em Phys. Lett.}  B}
\def\PRL{\em Phys. Rev. Lett.}
\def\PRD{{\em Phys. Rev.} D}

\def\be{\begin{equation}}
\def\ee{\end{equation}}
\def\bea{\begin{eqnarray}}
\def\eea{\end{eqnarray}}

\begin{document}

\title{INFLATION AND THE $B-L$ BREAKING SCALE}

\author{ R. JEANNEROT }

\address{Centre for Theoretical Physics, University of Sussex,\\ 
Falmer, Brighton, BN1 9QH, UK}


\maketitle\abstracts{
Inflation arises in supersymmetric grand unified theories (susy GUTs)
without fine tuning and
cosmic strings usually form at the end of inflation. Hence both strings and 
inflation contribute to the density perturbations in the very early universe
which lead to structure formation and to CMB anisotropies. This may give
us a hint as to the $B-L$ breaking scale. }
  
\section{Inflation and cosmic strings in susy GUTs}

Inflation requires that there was a period in the early universe when 
the vacuum energy density was non zero so that the cosmic scale factor 
grew exponentially. In susy theories, the scalar potential  
is the sum of F-terms and D-terms. Hence inflation 
can either
come from the non vanishing vev of a F-term or by that of a D-term. 
Inflation can also either come from the
visible sector or from the hidden sector which  breaks 
susy at low energy and communicates with 
the visible sector only via gravitational interaction. 
For a given GUT G, the full superpotential can 
formally be written as:
$W = W_{\rm GUT} (A_i) + W_{\rm infl}(S,\Phi,\overline{\Phi}) 
+ 
W_{\rm ew}(H_1,H_2) + W'_{\rm hidden}(B_i)$,
where $W_{\rm GUT}$ implements the breaking of G down to the 
$3_c \, 2_L \, 1_Y$ except from the breaking which is done in the inflaton 
sector
when inflation comes from the visible sector; 
$W_{\rm infl}$ leads to a period of inflation, $S$ is
a scalar singlet under G and plays the role of the inflaton, and when inflation
comes from the visible sector $\Phi$ and $\overline{\Phi}$ are Higgs superfields
which transform non trivially under G; $W_{\rm ew}$ breaks 
$3_c \, 2_L \, 1_Y$ down to $3_c \, 1_Q$;
$W'_{\rm hidden}$ makes the breaking in the hidden sector, except from
the part which is done in the inflaton sector when inflation comes from
the hidden sector.

The simplest superpotential for F-term inflation is given by~\cite{Fterm}
$W^F_{\rm infl} =\alpha S \Phi \overline{\Phi} - \mu^2 S$
and the scalar potential
$V^F_{\rm infl} = \alpha^2 |S|^2 ( |\Phi|^2 
+ |\overline{\Phi}|^2) +|\alpha \overline{\Phi} \Phi - \mu^2|^2 $,
$\alpha,\beta\geq 0$. This potential 
reduces the rank of G by one unit  and
leads to inflation. Setting chaotic initial conditions, the fields 
quickly settle down to the local minimum of the potential at
 $|S| > {\mu \over \sqrt{\alpha}} = S_c$ and
$\langle |\Phi |\rangle =  \langle | \overline{\Phi} | \rangle = 0$;
there is a non-vanishing 
vacuum energy density $V = \mu^4$, susy is broken, and inflation starts. 
Quantum corrections to the effective potential
help the inflaton to slowly roll down the potential. When $S$ falls below
$S_c$, inflation stops, the
fields settle down to the global minimum of the potential at
$\langle |\Phi | \rangle 
= \langle | \overline{\Phi} | \rangle = {\mu \over \sqrt{\alpha }}$
and $S = 0$, susy is restored
and the SSB associated with the vevs of $\Phi$ and 
$\overline{\Phi}$ takes place. It is easy to see that the above potential 
has got cosmic strings solutions, therefore cosmic 
strings form at the end of inflation.

The simplest superpotential for D-term inflation is given by~\cite{Dterm}
$W^D_{\rm infl} = \alpha S \Phi_{x} \overline{\Phi}_{x} $
and the scalar potential is
$V^D_{\rm infl} =\alpha^2 |S|^2 (|\Phi_x|^2 + |\overline{\Phi}_x|^2 ) +
\alpha^2 |\overline{\Phi}_x\Phi_x|^2 + {g^2 \over 2} (|\overline{\Phi}_x|^2
-|\Phi_x|^2 + \xi_x)^2 $,
we now assume the presence of a Fayet-Iliopoulos D-term $\xi_x$
which can only exist is G contains a $U(1)_x$ factor. 
The evolution of the fields is very similar to the
previous case. The local minimum is 
at $|S| > {g\over \alpha} \xi_x^{1\over 2}$,
 $\langle \Phi_x \rangle = \langle \overline{\Phi}_x \rangle = 0$  and the global 
susy minimum at $\langle S\rangle = \langle \overline\Phi \rangle = 0$,
$\langle |\Phi|\rangle = \xi_x^{1\over 2}$. The ${\rm U}(1)_x$ gauge symmetry
breaks at the end of inflation. It is easy to see that the scalar potential 
has also got cosmic strings solutions, 
and cosmic strings form at the end of inflation.

Hence in both types of scenarios cosmic strings form at the end of inflation,
both strings and inflation will contribute to the CMB anisotropies.

\section{CMB anisotropies and COBE normalisation}

It is common to expand the temperature fluctuations in the CMB in terms
of spherical harmonics: ${\delta T \over T }(\theta ,\phi)  = 
\sum_{l,m} a_{lm} Y_{lm}(\theta ,\phi)$,
and then work with the multipole moments
$C_l = {1 \over 2l+1} \sum_{m=-l}^{m=+l} |\! a_{lm}\! |^2$.
For a mixed scenario with inflation and cosmic strings the total $C_l$s
are given by
\begin{equation}
 C_l^{\rm tot} = C_l^{\rm infl} + C_l^{\rm str}
\hspace{.6cm} {\rm with} \hspace{.6cm}
 C_l^{\rm infl} \propto \delta_H^2 \hspace{.6cm} {\rm and}\hspace{.6cm}
 C_l^{\rm str} \propto (G\mu )^2 \label{eq:Clmix}
\end{equation}
where, $\delta_H^2$ is the spectrum of density perturbations from inflation
at horizon crossing and $\mu$ is the string mass 
per unit length; 
they can be calculated exactly from the slow roll parameters.
In the case of D-term inflation we get
$G\mu = {2 \pi}\left( {\xi_x \over M_{pl}^2}\right)$
and $\delta_H = {\alpha_{60} 256 \pi^2 121 \over 75}
\left({\xi_x \over M_{pl}^2}\right)$, where $\alpha_{60} \simeq 1$.

Combining Eqs.~\ref{eq:Clmix} we get the normalisation
equation for a mixed scenario with inflation and cosmic strings:
$1 = \left({G\mu \over (G\mu)^{\rm norm}}\right)^2 + 
\left({\delta_H \over \delta_H^{\rm norm}} 
\right)^2 .$
Using the normalisation to COBE for strings from~\cite{Shel} and that for 
inflation from~\cite{Lid}, we get $(G\mu)^{\rm norm} = 1.05 \times 10^{-6}$
and $\delta_H^{\rm norm} =  1.94\times 10^{-5}$; we then find that the 
${\rm U}(1)_x$ SSB scale is constraint by COBE to be
$\xi_x^{1\over 2} = 4.7 \times 10^{15}$ GeV, and we also find that cosmic 
strings can contribute to the $C_l$s up to the level of 75\%.

\section{Application: models with an intermediate $U(1)_{B-L}$}

We now consider the general SSB pattern
\begin{eqnarray*}
G\times{\rm susy} &\rightarrow& ... 
 \stackrel{M_G}\rightarrow 3_c \, 2_L \, 2_R \, 1_{B-L} \times{\rm susy}\\
& \stackrel{M_R}\rightarrow &3_c \, 2_L \, 1_R \, 1_{B-L} \times{\rm susy}
\stackrel{M_{B-L}}\rightarrow 3_c \, 2_L \, 1_Y \, Z_2 \times{\rm susy}
\end{eqnarray*}
where the discrete $Z_2$ is a subgroup
of $U(1)_{B-L}$ and plays the role of R-parity, and susy is broken 
at $\sim 10^{3}$ GeV.
At $M_G$ 
topologically stable monopoles in contradiction with observations form; at 
$M_R$ more stable monopoles form; at $M_{B-L}$ $B-L$ cosmic strings
(the Higgs field forming the string is the same Higgs field used to break 
$B-L$) form, they are good candidate for baryogenesis. 
To satisfy observations, a period of inflation is needed between $M_R$ 
and $M_{B-L}$. We can also consider models without the intermediate $M_G$
or $M_R$, the conclusion upon $M_{B-L}$ would be the same.

If inflation comes from the GUT itself, the superpotential in the inflation 
sector is $W_{\rm infl}(S,\overline{\Phi}_{B-L}, \Phi_{B-L})$, $\overline{\Phi}_{B-L}$
and $\Phi_{B-L}$ are Higgs used to break $B-L$ (if ${\rm G} = {\rm SO}(10)$ 
they 
would be a pair of $126 + \overline{126}$-dimensional representations). In
this case, COBE constraints the $B-L$ breaking scale to be $M_{B-L} = 4.7 
\times 10^{15}$ GeV and $B-L$ cosmic strings form at the end of inflation.

If inflation comes from the hidden sector, then there is
considerable freedom in choosing $M_{B-L}$. However, both $M_G$ and $M_R$
must be greater than $4.7 \times 10^{15}$ GeV for the monopoles problem to be 
solved, unless $M_G, M_R \ll 10^{11}$ GeV. In that case, the strings which form 
at the of inflation only communicate to the visible sector via gravitational 
interactions. 

The above class of models is phenomenologically very interesting, it 
provides both CDM and HDM in the form of the LSP and a massive neutrino
respectively, baryogenesis via leptogenesis takes place at the end of inflation,
and both strings and inflation must be at the origin of density perturbations in the early
universe which lead to structure formation and CMB anisotropies.

\section*{Acknowledgments}
This work was supported by PPARC grant no GR/K55967.

\end{document}